\documentclass[final,5p,times,twocolumn]{revtex4-2}

\usepackage{graphicx}
\usepackage{dcolumn}
\usepackage{bm}
\usepackage{color}
\usepackage{amssymb}
\usepackage{amsthm}
\usepackage{setspace}
\usepackage{amsmath}
\usepackage[hidelinks,colorlinks = true,
linkcolor = black,
urlcolor  = black,
citecolor = black,
anchorcolor = black]{hyperref}
\usepackage{ulem}
\usepackage{comment}
\usepackage [latin1]{inputenc}
\usepackage{float}

\newcommand{\be}{\begin{equation}}
\newcommand{\ee}{\end{equation}}
\newcommand{\ba}{\begin{eqnarray}}
\newcommand{\ea}{\end{eqnarray}}
\newcommand{\bd}{\begin{displaymath}}
\newcommand{\ed}{\end{displaymath}}
\newcommand{\bea}{\begin{eqnarray}}
\newcommand{\eea}{\end{eqnarray}}

\newcommand{\beq}{\begin{equation}}
\newcommand{\beqar}{\begin{eqnarray}}
\newcommand{\eeq}[1]{\label{#1} \end{equation}}
\newcommand{\eeqar}[1]{\label{#1} \end{eqnarray}}

\begin{document}


\title{{\color{black}
PIC simulations of laser-induced proton acceleration\\
by resonant nanoantennas for fusion}
}

\author{
Istv\'an Papp $^{1,2,11}$,
Larissa Bravina $^4$,
M\'aria Csete  $^{1,5}$,
Archana Kumari $^{1,2}$,
Igor N. Mishustin  $^{6}$, \\
Anton Motornenko   $^6$, 
P\'eter R\'acz $^{1,2}$,
Leonid M. Satarov  $^{6}$,
Horst St\"ocker  $^{6,7,8}$,
Andr\'as Szenes  $^{1,5}$,
D\'avid Vass   $^{1,5}$,
Tam\'as S. Bir\'o $^{1,2}$,
L\'aszl\'o P. Csernai $^{1,2,3,9}$,
Norbert Kro\'o $^{1,2,10}$\\
(part of NAPLIFE Collaboration)\\}
\bigskip

\address{
$^{1}$ Wigner Research Centre for Physics, Budapest, Hungary\\
$^2$ Hungarian Bureau for Research Development and Innovation \\
$^3$ Dept. of Physics and Technology, University of Bergen, 5007 Bergen, Norway\\
$^4$ Department of Physics, University of Oslo, Norway\\
$^5$ Dept. of Optics and Quantum Electronics, Univ. of Szeged, Hungary\\
$^6$ Frankfurt Institute for Advanced Studies, 60438 Frankfurt/Main, Germany\\
$^7$ Inst. f\"ur Theoretische Physik, Goethe Universit\"at Frankfurt, 
60438 Frankfurt/Main, Germany\\
$^{8}$ GSI Helmholtzzentrum f\"ur Schwerionenforschung GmbH, 
64291 Darmstadt, Germany\\
$^{9}$ Csernai Consult Bergen, Bergen, Norway \\
$^{10}$ Hungarian Academy of Sciences, 1051 Budapest, Hungary\\
$^{11}$ HUN-REN Centre for Energy Research, Budapest, Hungary
}

\begin{abstract}
Rapid recent development in laser technology and methods learned from 
relativistic heavy ion physics led to new possibilities for fusion. 
Using a Hydrogen rich UDMA-TEGDMA polymer fusion target, laser 
irradiation ionizes the target. If we implant nanoantennas into 
the target resonating to the laser light frequency massive number of 
electrons of the ionized plasma resonate within the nanoantenna forming 
a so called nanoplasmonic wave. Our kinetic model simulation with a 
Hydrogen target indicates that the field of these resonating electrons 
attracts and accelerates the surrounding protons of the plasma to 
multi-MeV energy. These protons are then energetic enough to achieve
nuclear transmutation and fusion reactions. Without resonating nanoantenna 
there is no such collective proton acceleration, no energetic protons, 
and nuclear reactions at 30 mJ laser pulse energy.
\end{abstract}

	

\maketitle
\section{INTRODUCTION}

In recent years, NAnoPlasmonic Laser Inertial Fusion Experiments (NAPLIFE)  
\cite{CsEA2020} 
were proposed to achieve in an improved way nuclear fusion 
as a clean and efficient source of energy. The aim is to achieve laser driven 
fusion in a non-thermal collider configuration to avoid instabilities during 
ignition. Simultaneous volume (or "time-like") ignition 
\cite{Cs1987,CS2015} 
can be achieved with enhanced energy absorption 
with the help of nanoantennas implanted into the target material
\cite{CsKP2018}. 
This should prevent the development of 
the mechanical Rayleigh-Taylor 
instability. Furthermore, the nuclear burning should not propagate
from a central hot spot to the outside edge as the ignition is
simultaneous in the whole target volume.

However, before this approach can be put into practice, it is crucial to 
understand the resilience of these nanoantennas under various conditions. 
To this end, recent studies have been conducted to investigate the behavior 
of the nanoantennas in vacuum 
\cite{Frontiers2023} 
as well as in a UDMA-TEGDMA medium. These studies provide valuable insights 
into the behavior of the nanoantennas and their potential as a key element 
in achieving laser induced fusion with simultaneous volume ignition
\cite{CsMi-add}.

Non-equilibrium and linear colliding configuration 
have been introduced already 
\cite{Barbarino2015,Bonasera2019}.
Here we study the idea of layered flat target fuel with
embedded nanoantennas, that regulate laser light 
absorption to enforce simultaneous ignition.
We plan a seven-layer flat target with different
nanorod densities. 
\cite{CseteX2022,BA2020,OMEX}.
To prepare such a layered target, the ignition fuel
(e.g. deuterium, D, tritium, T, or other nuclei for fusion)
is embedded into a hard thin polymer material of seven,
3$\mu$m thick layers. These 
polymers are
Urethane Dimethacrylate  (UDMA) and 
Triethylene glycol dimethacrylate (TEGDMA) in (3:1) mass ratio 
\cite{UDMA1,BA2022}.

The linear colliding beam configuration allows for a much simpler flat 
target configuration
\cite{CsEA2020,BA2020}, 
utilizing a non-equilibrium, ignition dynamics. 
This  is realized by  relativistic collisions of two
target slabs, produced by using the Laser Wake Field Acceleration 
(LWFA) mechanism
\cite{P-EA2021}.
The use of femtosecond laser pulses leads to rapid ignition with high, 
beam-directed collision velocities.

The benefits of non-equilibrium configuration were first introduced  in ref.
\cite{Barbarino2015}, 
and experimentally tested in linear colliding configuration
\cite{Bonasera2019}.
Similar pre-compression was reached by this configuration as in the
National Ignition Facility (NIF). 

{\color{black}
Proton acceleration by laser irradiation is discussed  in the 
literature for more almost two decades 
\cite{Pukhov1996,Pukhov1999,Snavely2000,Maksimchuket2000}, 
almost exclusively in the Target Normal Sheath Acceleration 
(TNSA) configuration, i.e. shooting the laser beam across a thin 
target foil. In several of these works, nanotechnology was 
considered as a layer or a hole in the layers of the target foil 
\cite{Cristoforetti2020,Leonida2021,Cantono2021,Mehrangiz2022}. 
None of these works mentioned embedded, resonant nanoantennas in 
the target material. In contrast to the overwhelming thermal 
considerations in these previous works, our approach takes advantage 
in direct, non-thermal reactions also in the transformation of 
laser energy into the required higher nuclear energy level protons.
}
\section{Initial Studies}

\begin{figure}[!htb]  
\begin{center}
\includegraphics[width=0.99\columnwidth]{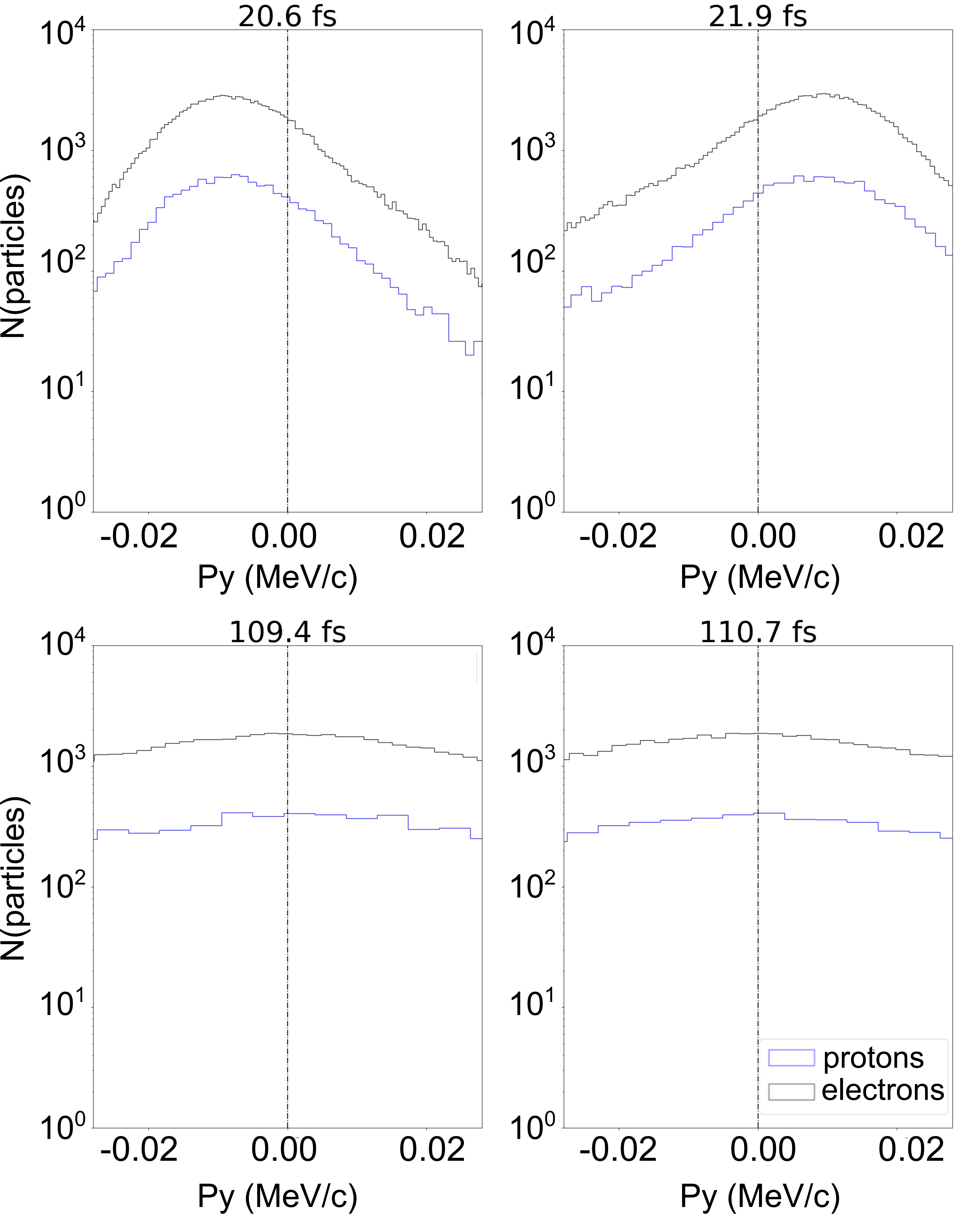}
\end{center}
\caption{
(color online)
A snapshot of pairs of particle number 
oscillations half of a period (T/2) apart are
shown at an earlier (top) and later (bottom) stage.
The histogram of the $y$ directed momentum distribution is shown. 
Black lines show momentum distribution of the electron 
marker particles, while black lines
show protons surrounding the nanorod. The vertical line in the middle separates 
particles going up (right) and down (left) along the nanoantenna. 
The highest peak of the proton momentum distribution is not in total 
anti-phase of the highest peak of the electrons, they follow the electrons
with a small delay. This suggests that the protons are moved not by the 
electric field of the laser beam but in the near field of 
the Localized Surface Plasmon Polaritons.
The widening of the momentum distribution can be attributed to the kinetic 
model assumption, where both the conduction and binding electrons are 
considered freely movable. The resonance of the nanorod has a period of
$T=2.65$ fs.}
\label{Fig-1}
\end{figure}

In our previous studies we have demonstrated that
in addition to achieving simultaneous energy deposition
there are other significant effects at laser intensities of 
I = $4 \times 10^{15}$ W/cm$^2$, I = $4 \times 10^{17}$ W/cm$^2$:
(i) the collective motion of conduction electrons enhances the electric 
field while contributing to the absorption of laser light 
\cite{CseteX2022}; 
(ii) the high number of oscillating electrons in the nanorod
\cite{PRX-E2022}
accelerate 
those protons, which are close to the surface of the nanorod, 
Fig. \ref{Fig-1}; 
(iii) the high density electron screening leads to 
lowering the Coulomb barrier, the phenomenon presents in stars as well 
\cite{Aliotta2022}. 

\begin{figure}[H]  
\begin{center}
\includegraphics[width=0.90\columnwidth]{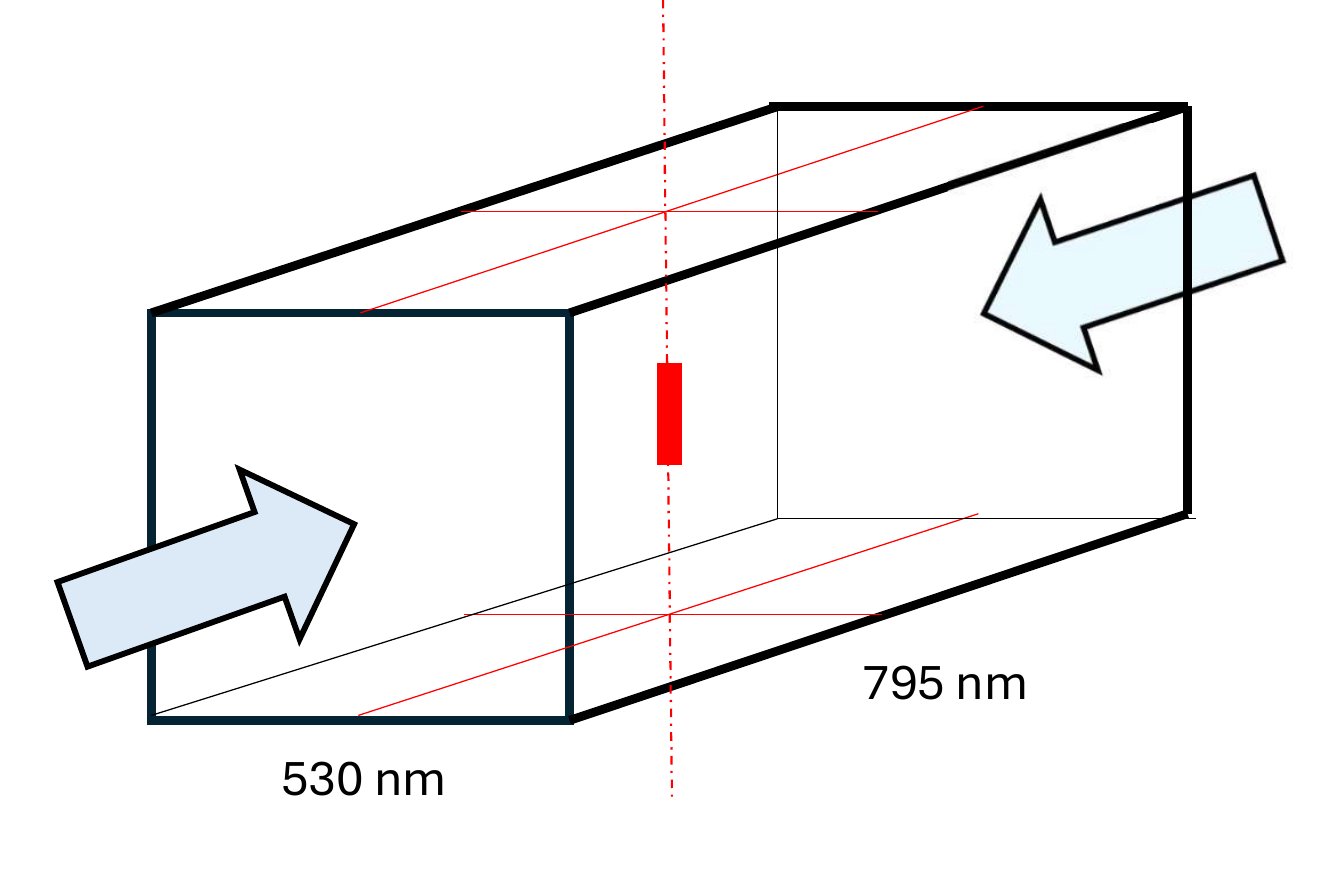}
\end{center}
\caption{(color online) 
{\color{black}
Schematic view of the calculation box (CB) and the nanorod orientation 
relative to the two-sided laser irradiation. The CB is divided into 5 nm
qubic cells in the PIC method. Lagrangian fluid cells of different particle
species are represented by marker particles.}}
\label{Fig-2}
\end{figure}

Let us consider resonant nanoantennas embedded in 
Hydrogen-rich polymer and/or with Hydrogen-rich coating. 
We assume that the antennas are orthogonal to the direction 
of laser irradiation
and are made of good plasmonic material with negative refraction 
index such as gold, silver, or aluminum. 
Linear irradiation from two opposite directions, with electric
fields in phase is envisaged because in this case no net momentum 
is transferred to the
nanorods, but full energy is deposited to the antenna \ref{Fig-2}. 
The antenna should have the
proper effective length in the given target and antenna material,
as well as the length/diameter aspect ratio 
\cite{Nov2007,eps-inf}.
In this case the valence electrons will oscillate between the two ends 
of the nanorod
due to the fluctuating in-phase electric field from the two laser 
beams. In case of
sufficient intensity ($ \approx 10^{17}$ W/cm$^2$) of the 
laser beams the irradiated
material will be strongly ionized, and its molecules will be 
dissociated into single atoms.
(If the laser beam intensity is weaker molecules will fragment 
into groups and single
atoms.) High Hydrogen content target materials or nanorod coatings will 
provide freely movable
protons, which will be accelerated due to the correlated 
motion of electrons and
protons caused by the electric field generated on the nanoantennas. 
The large
number of electrons in the nanoantennas (of length 
$\lambda_{eff}$ = 75-85 nm)
produce enormous field strength ($E_y = 3\times10^{12}$ V/m), 
\cite{PRX-E2022}, 
which may provide
sufficient energy for the smaller number of nearly 2000 times 
heavier protons.
      
This acceleration mechanism is like high energy particle accelerators 
where several
electrode pairs with fluctuating electric charge accelerate the protons 
(e.g., in LHC in
one pair of electrodes (cavity) the charges are about 10 cm apart and 
the potential is
about $10^6$ V/m. This is then repeated many times in a length of 27 km). 

One nanorod corresponds to one electrode pair in the particle accelerators. 
However, the
electric field strength is six orders of magnitude higher, and the 
length of acceleration
is six orders of magnitude shorter. These estimates are based on the 
EPOCH PIC kinetic model 
\cite{PRX-E2022,Frontiers2023}.
The number of electrons is decisive for this mechanism. Here quantum 
effects play an important role 
\cite{Vysotskii2013}. 

\section{{Description of the Model and Simulations}}

Plasmons on such nanoantennas are usually studied theoretically by solving 
Maxwell's equations in simulations involving 
finite difference
time-domain methods. These approaches are efficient for computers, 
however, some of 
the important phenomena are either neglected this way or 
included typically by 
extra fitting parameters. 
The motion of free electrons is only considered indirectly via the bulk 
permittivity of the metal. We use classical 
dielectric function of the free electron gas: 
$\epsilon (\omega) = 1 - \omega_{p}^2 /( \omega^{2} 
+ i \gamma \omega)$ with $\omega_{p}$ 
being the plasma frequency  
\cite{Maier2007}.
The electron-electron interactions are partly included
by using an effective mass for the electrons.

In the above formula  $\gamma$ is the
collision frequency, providing the damping effect.
In our simulations we use the 
electron number density, $n_e$, 
to randomly distribute electron-like marker particles on the metal surface.
In our simulations we apply the Particle-in-Cell (PIC) method
\cite{Harlow}, 
where {marker-particles} (representing large number of real particles)
move  in continuous phase space, whereas densities and 
currents are computed in stationary mesh cells.
We use the EPOCH multi-component PIC code by T. D. Arber et al.
\cite{Arber2015}. 
This approach proved to be efficient for analyzing 
the electron dynamics in different plasmon modes, and modeling electron 
spill-out effects 
\cite{Ding2020}.

\begin{figure}[H]  
\begin{center}
\includegraphics[width=0.90\columnwidth]{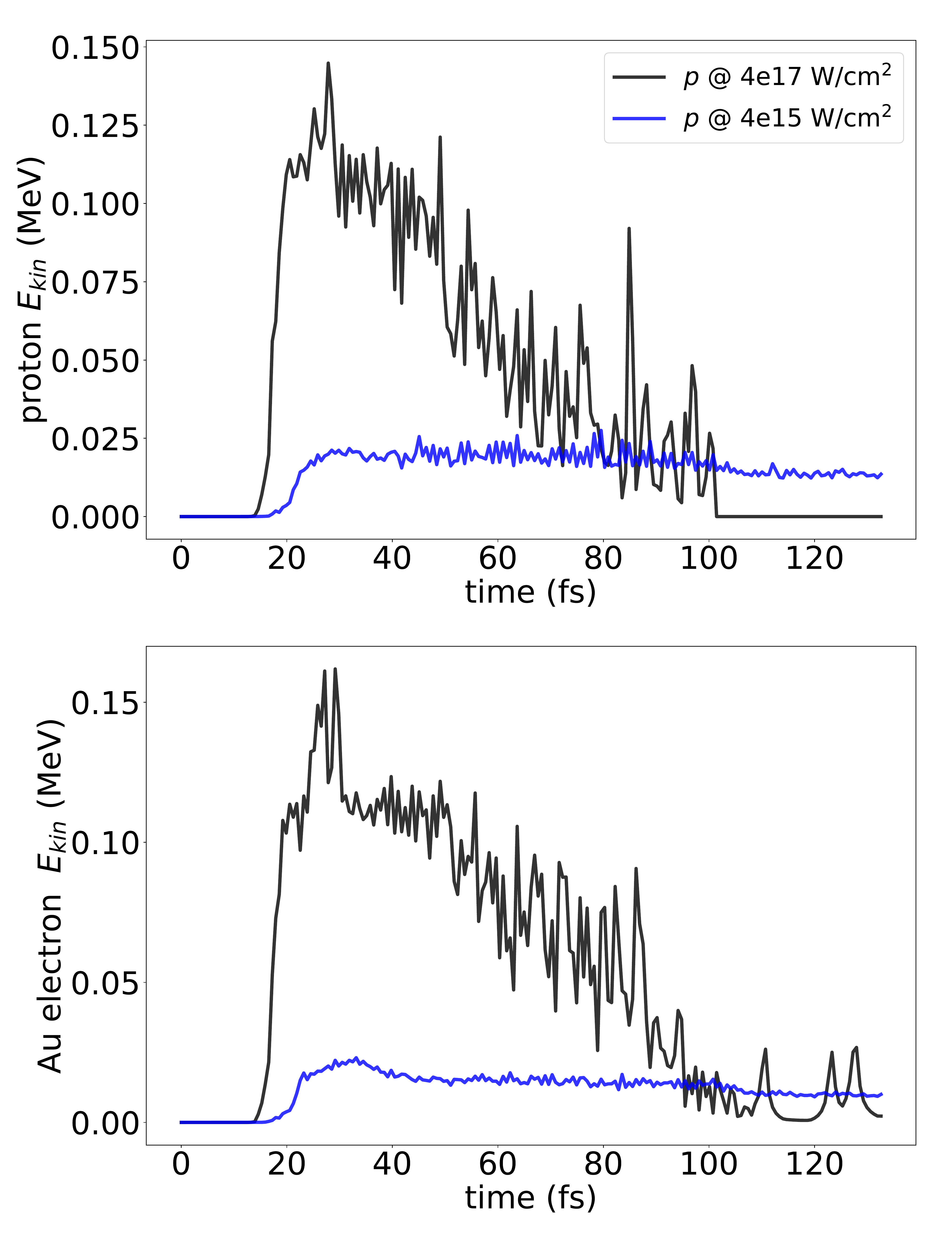}
\end{center}
\caption{(color online) 
{\color{black}
Here we show the derived average particle kinetic energy calculated 
for each cell for the individual macro particles.
We show the maximum of that energy at each time. This result closely
correlates with the experimental results found at Wigner Research 
Center where the intensities are
roughly at 4$\cdot 10^{17}$W/cm$^2$. 
\cite{KrooArXiv,Szokol,RigoArXiv}}}
\label{Fig-3}
\end{figure}

\begin{figure}[b]  
\begin{center}
\includegraphics[width=0.99\columnwidth]{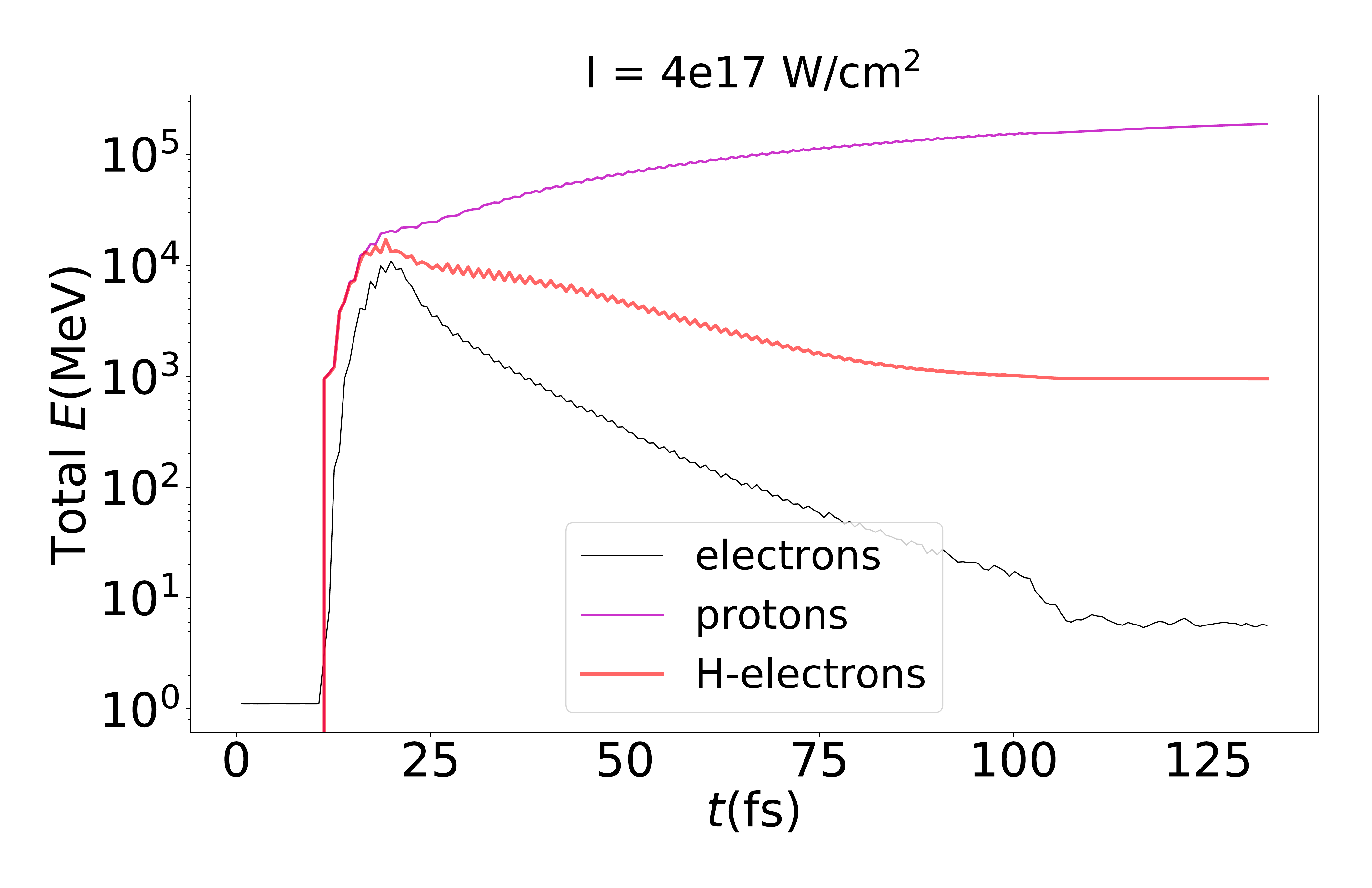}
\end{center}
\caption{
(color online)
We consider a laser pulse of intensity 
$I= 4 \cdot 10^{17}$W/cm$^2$ and duration of $106$fs.
Here we show the time dependence of the total $y$ directed 
energy of the conducting electrons in the nanorod (black line), 
protons resulting from ionization (magenta line)
and Hydrogen electrons  remaining from ionization (red line).
It is shown that at 10 fs the laser light starts ionizing the 
Hydrogen atoms, while at the same 
time the Hydrogen-electrons follow the collective motion of 
the gold nanorod
electrons. The protons follow the same periodic motion with a 
phase delay.}
\label{Fig-5}
\end{figure}

\begin{figure}[t] 
\begin{center}
\includegraphics[width=\columnwidth]{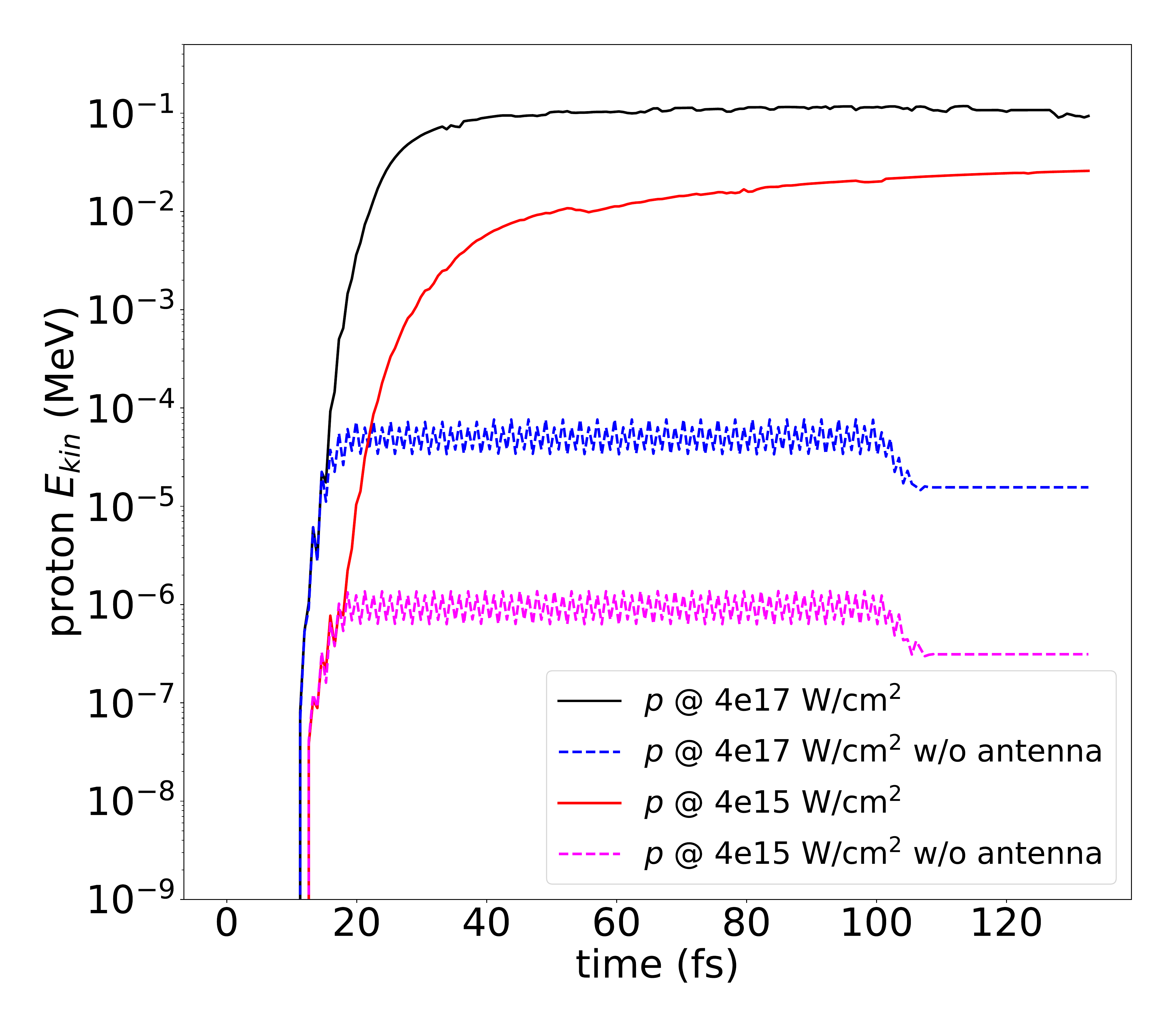} 
\end{center}
\caption{{\color{black}
(color online) 
Here we show the time evolution of the derived average 
kinetic energy of the proton marker particles. 	 Simulated at intensities 
of 4$\cdot 10^{15}$W/cm$^2$ and 4$\cdot 10^{17}$W/cm$^2$.
We show the results of simulations in the Hydrogen medium with 
and without the nanoantenna.
In the non-doped case, for both intensities a plateau is reached, 
which lowers at the end of the shot, when no more acceleration can be 
achieved. However, when nanorod is present, the plasmonic electron 
bunch's motion further accelerates the protons by to three orders 
of magnitude.}}
\label{Fig-8}  
\end{figure}

\begin{figure}[H]  
\begin{center}
\includegraphics[width=\textwidth]{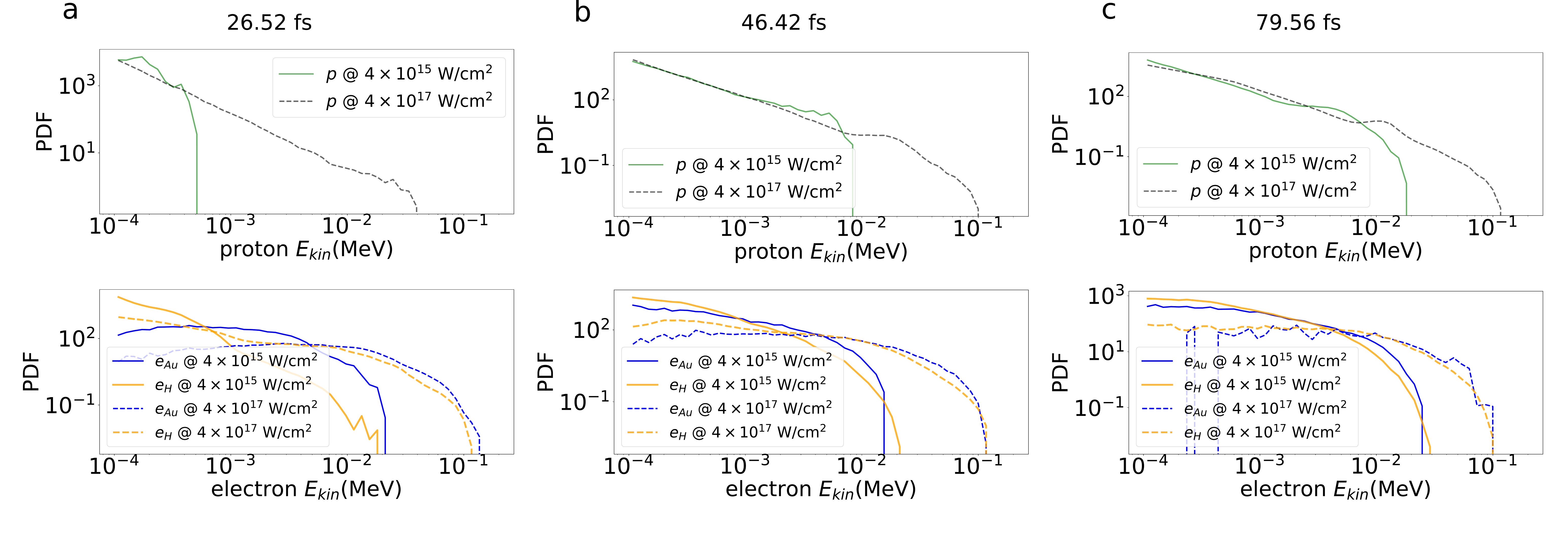}  
\end{center}
\caption{(color online) 
{\color{black}
Probability Density Function (PDF) normalized to 
unity for average particle energy
for ionized Hydrogen, the freed electrons and gold plasmonic 
electrons. Simulated at intensities of
4$\cdot 10^{15}$W/cm$^2$ and 4$\cdot 10^{17}$W/cm$^2$, and for different 
irradiation times (a,b,c). For the 10-100 keV kinetic energy domain 
it increases significantly with time and laser beam intensity.}}
\label{Fig-6} 
\end{figure}

{\color{black}
Two simulations were performed:\\
(1)  The electron screening effect is examined on a thin layer of
proton "coating" surrounding the resonant nanoantenna with density of 
$6.5\times 10^{28}$ proton/m$^3$. 
In the experiments the nanorod has a Dodecanethiol (DDT) capping: 
CH$_3$(CH$_2$)$_{11}$SH.
This molecule is 1.3 nm long, with a molar mass 202.39 g/mol and density 
of 0.845 g/cm$^3$.
The thickness of the layer covering the nanorod is smaller than this, as 
the molecules bind to the surface at a certain angle. The attachment of 
these self-organized molecular monolayers to gold surfaces have been 
studied by many authors, 
and the literature has a wide  range of data on coverage 
and angles for various cases 
\cite{LarsM2005}.
The densest conformation only occurs on a 
perfectly (111)-oriented surface (111 here refers to the plane of atoms 
in a crystal described by the Miller indices 
\cite{Ashcroft1976}). 
This was 
not taken into account; the resolution of the simulation box was greater 
than the size of this molecule. 
Results can be seen in Figures \ref{Fig-1} and \ref{Fig-2}.

(2) The  nanoantenna was inserted in a box filled by neutral Hydrogen atoms 
with density of  $4.27\times 10^{27}$ atoms/m$^3$ (close to liquid density), 
in the limit where we can still treat every marker particle species 
participating as a plasma component with kinetic approach.
The Hydrogen atom consisted of 3 particle species: neutral Hydrogen atom, 
which could be ionized with the laser or by collisional ionization 
\cite{Perez2012}, 
protons and electrons resulting from the ionization. 
Furthermore, the gold nanorods consisted of 2 species: stationary 
Au$^{3+}$ ions, and the moving conduction electrons. 
Results of these simulations can be seen in figures 
\ref{Fig-4}-\ref{Fig-6}.

Since the nanorod in our model calculation is 85 nm long
and its diameter is $D=25$ nm it has many differently
oriented surfaces. Thus, we may only estimate the average. 
We considered 3 electrons in the conduction band.
Irradiation was done by 
$4 \times 10^{15}$ W/cm$^2$ (results shown on 
Figures \ref{Fig-2}, \ref{Fig-3}, \ref{Fig-5}, \ref{Fig-6}) 
and $4 \times 10^{17}$ W/cm$^2$ intensity (results shown on 
Figures \ref{Fig-4}, \ref{Fig-5}, \ref{Fig-6})
plane wave, at 795 nm   
wavelength, and pulse duration of 106 fs, just like in previous works 
\cite{PRX-E2022,Frontiers2023}. 
Similarly, we were irradiating 
a calculation box (CB {\color{black} see Figure \ref{Fig-2}}) of cross section 
$S_{CB} = 530 \cdot  530$nm$^2$ = $2.81 \cdot 10^{-9}$ cm$^2$  
and of length
$L_{CB} = \lambda =795$ nm. 
The volume of CB being 0.2233 $\mu$m$^3$, the total volume of H would be 
$V_{CB} - V_{nanorod}$, 
where 
$V_{nanorod} = 85 \cdot \pi*(D/2)^2 = 41724.27$ nm$^3$. 

In the PIC method marker particles corresponding to
Lagrangian fluid cells moving together with the flow
of a given type of particle (in our case Hydrogen atoms earlier, Hydrogen1 ions and their electrons $e_H$, Au 3+ ions, and the plasmonic electrons of the gold)
}

\section{Results}

\begin{figure}[t] 
\begin{center}
\includegraphics[width=\columnwidth]{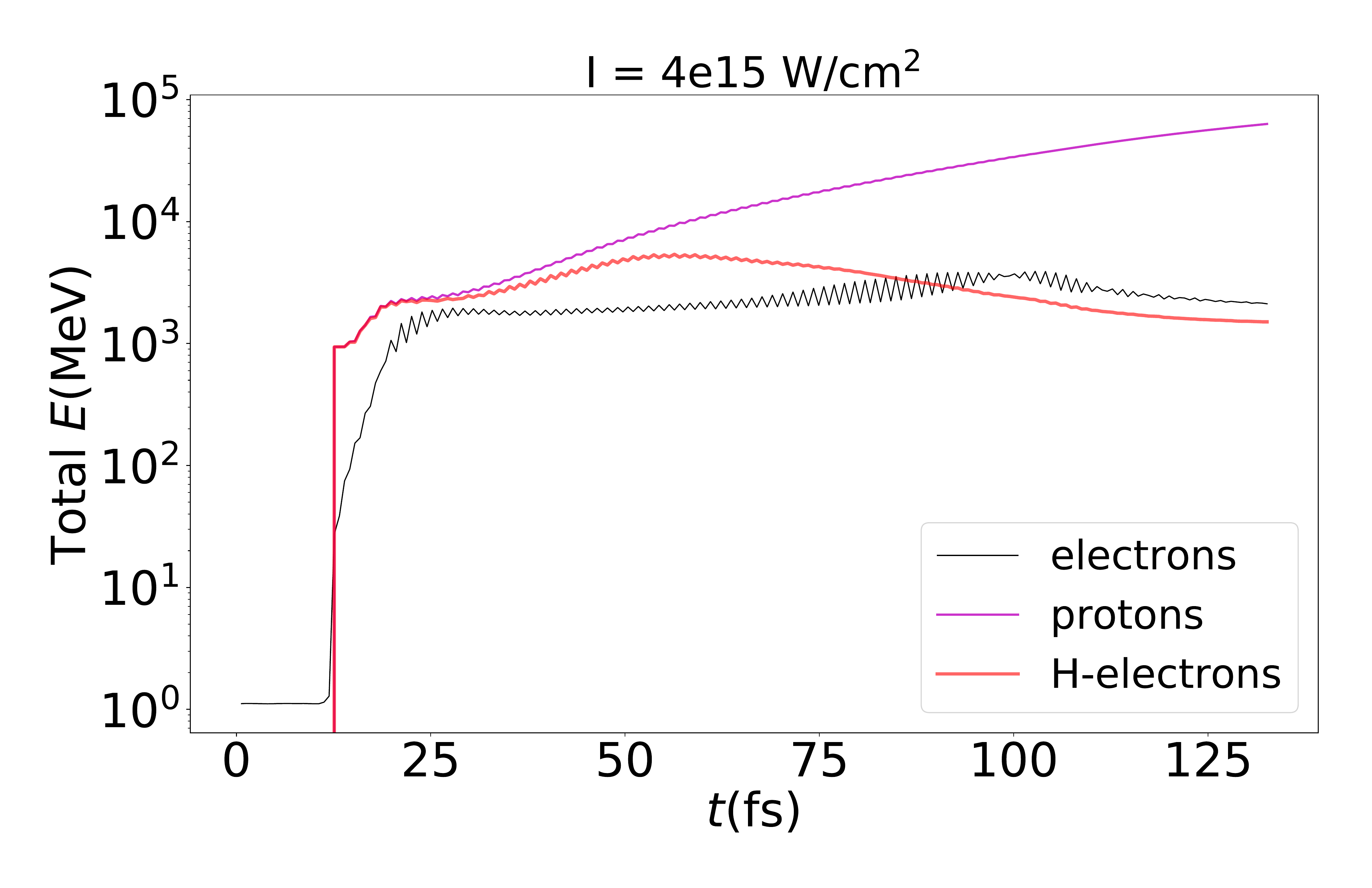}
\end{center}
\caption{
(color online)
We consider a laser pulse of intensity 
$I= 4\cdot 10^{15}$W/cm$^2$ and duration of $106$fs.
Here we show the time dependence of the total $y$ directed 
energy of the conducting electrons in the nanorod (black line) ,
protons resulting from ionization (magenta line)
and Hydrogen electrons  remaining from ionization (red line).
It is shown that at 10 fs the laser light starts ionizing the Hydrogen atoms,
while at the same time the Hydrogen-electrons follow the collective motion of 
the gold nanorod electrons. The protons follow the same periodic motion 
with a phase delay.}
\label{Fig-4}
\end{figure} 

As the Hydrogen atoms are ionized, the number of neutral H marker particles 
in the simulation box decreases 
and proton and  electron marker particles take their places. The leaving 
electrons from 
the ionization in the close vicinity, coupled to the plasmonic electrons 
through the laser, enhance further the  {\color{black}
accelerating effect on the protons at both laser intensities, see 
Figs. \ref{Fig-3}, \ref{Fig-4}.
This effect further increases the proton's energy as seen in 
Fig. \ref{Fig-3}.}

As one can see in Fig. \ref{Fig-3}, the electro-magnetic field drives the 
conduction electrons into resonant fluctuations. In case of long irradiation
times, $T_{pulse} \gg T$, our model describes a stationary configuration, 
adequate for gentle acceleration and compression. This would precede the 
short and extremely energetic ignition pulse in a Laser Wake Field 
Collider configuration
\cite{P-EA2021}. 
At the transition from gentle irradiation to a few fs short ignition
the nanoantennas must not be
destroyed. The solid-state structure of the nanoantennas would 
be destroyed  when 
a significant part of the electrons left the nanoantennas.
This process takes a few ns, sufficient for the
amplified ignition
\cite{P-EA2021}.
It is also worth mentioning that net (positive) electric charge generated
in nanorods should induce strong electric fields around the rods. These
fields may be strong enough to accelerate protons in the UDMA material
outside the rods. As demonstrated in ref. 
\cite{CsMi-add}, the accelerated proton
energies may reach 10-20 MeV or even more, depending on the laser
intensity. Such proton energies are sufficient to induce endothermic
nuclear reactions like p+13C$\rightarrow$ d+12C and 
p+12C$\rightarrow$d+11C. Preliminary
results on laser-induced deuteron production are reported in ref. 
\cite{RigoArXiv}.

Note that for in-medium simulations 
the speed of light, as well as the wave lengths are reduced by the 
refractive index to
$c^* = c/n$ and $\lambda^* = \lambda/n$. 

Consequently, in a dielectric medium
like Urethane Dimethacrylate (UDMA) the resonant antenna length is reduced
by $n=1.5$ also, and further reduced by the
"thick" aspect ratio (25:75) cf. ref.
\cite{CseteX2022},
(see also the introductory comments based on ref. 
\cite{Nov2007}.)

It is instructive to compare two acceleration methods.
In the classical particle accelerators like the LHC, 
the potential difference between the neighboring RF cavities is 
$\sim 5 \cdot 10^6$ V/m, but between the two ends of the resonant
nanoantennas this is $\sim 2 \cdot 10^{12} - \cdot 10^{13}$ V/m. 
\cite{P-EA2021,PRX-E2022,Frontiers2023}.

At the LHC 
this field accelerates protons between the RF cavities on a $\sim 12.5$ cm
length, while by using resonant nanoantennas the length of acceleration
is  ~ 70 - 100 nm. At LHC the beam acceleration takes about 20 minutes
and a million rotation in 27 km length, we have a ~ 20 - 30 $\mu$m target
thickness and 50-100 fs laser pulse length.
Of course, at LHC one accelerates about $10^{11}$ protons per bunch,
while at the presently used laser facility at Wigner RCP (WRCP) with maximal 
laser pulse energy of ~ 30 mJ.

\onecolumngrid

\twocolumngrid

\section{{Conclusions}}
\label{DNR}

At the high contrast SYLOS laser facility, ELI-ALPS Szeged, Hungary, 
with higher laser
beam intensity we expect to have protons at higher MeV energies 
and/or significantly larger proton numbers than at WRCP.

The model presented in this paper is idealized but gives a possibility 
for us to capture the behavior of plasmonic nanorods on a small scale.
Most importantly this kinetic model simulation has demonstrated that 
the collective effect of many electrons can provide
a coherent proton acceleration. With proper target configuration
this effect can be further enhanced. 

{\color{black}
The mechanism of the accelerating effect of the plasmonic electrons is 
more visible at smaller
intensity (Fig. \ref{Fig-3}) due to less electron spill-outs leading 
to longer plasmon lifetime.
The Hydrogen gets ionized, and the protons and electrons get accelerated 
after the collective motion
is reached in the near field enhancement. This can be observed on 
Fig. \ref{Fig-4} with the transition from the
 plateau to increase in the average energy. On Fig. \ref{Fig-4} such
transition is not observed due to faster 
response from the conducting electrons reaching higher energies,  
dragging the protons, and shortening plasmonic lifetime (this effect 
is more noticeable on \ref{Fig-6}).}

\begin{figure}[H]  
\begin{center}
\includegraphics[width=0.90\columnwidth]{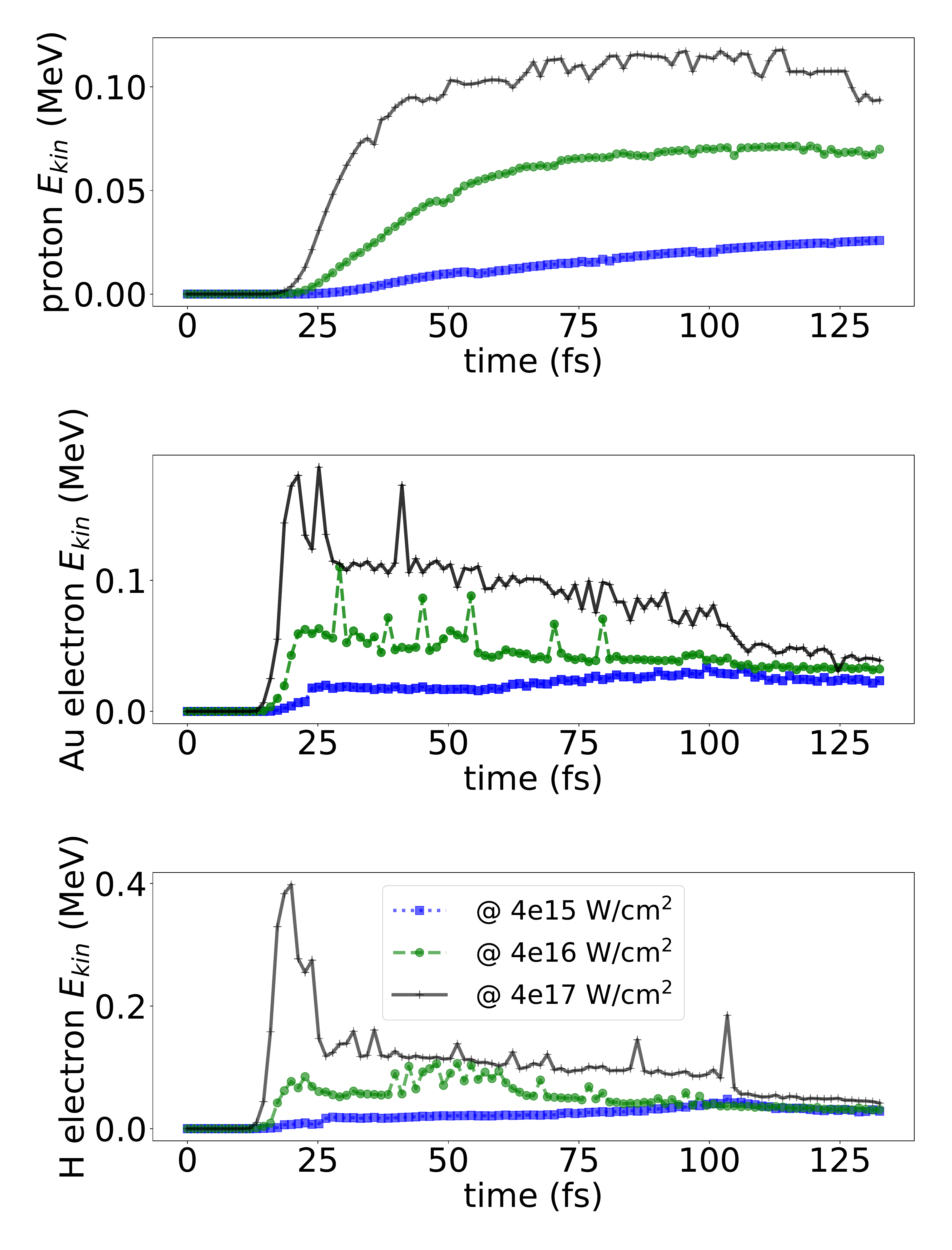}
\end{center}
\caption{(color online) {\color{black}
Here we show the time evolution of the derived average particle energy 
calculated in each cell for different macroparticle species, we plot 
the maximum at each time step.
Simulations were performed at different intensities:		
4$\cdot10^{15}$W/cm$^2$ (blue),  4$\cdot10^{16}$W/cm$^2$ (green), 
4$\cdot10^{17}$W/cm$^2$ (black)}.
}
\label{Fig-7}   
\end{figure}

There are numerous theoretical
predictions for Low Energy Nuclear Reactions. However, first we must
experimentally verify how these phenomena could be realized at higher
energies, and what reaction rates can be achieved. 
{\color{black} 
Most of the experimental realizations of the ideas from the field of 
Low Energy Nuclear Reactions lead to very low fusion reaction rates, 
which are not 
sufficient for massive industrial realization of these ideas. 
So far, no experimental 
results have shown the opposite. Our high energy non-thermal 
special space-time dynamics 
in principle can be scaled up to industrial levels. This is 
supported by the known 
NIF results also. We expect to have observed experimental rates soon, 
the first 
validation experiments are in progress and are under publication 
\cite{RigoArXiv,KrooArXiv,Szokol}.}

Our PIC simulations indicate that the angular distribution of proton
and ion emission in two sided irradiation becomes more peaked with time,
which is the major contribution to increased particle emission energy
\cite{HBT1,HBT2}.

\section*{Acknowledgments}
Enlightening discussions with Johann Rafelski are gratefully acknowledged.
Horst St\"ocker acknowledges the Judah M. Eisenberg Professor Laureatus 
chair at Fachbereich Physik of Goethe Universit\"at Frankfurt.
L\'aszl\'o P. Csernai acknowledges support from
Wigner Research Center for Physics, Budapest (2022-2.1.1-NL-2022-00002).
We would like to thank the Wigner GPU Laboratory at the Wigner Research Center 
for Physics for providing support in computational resources.
This work is supported in part by the 
Frankfurt Institute for Advanced Studies, Germany,
the Hungarian Research Network,
the Research Council of Norway, grant no. 255253, and
the National Research, Development and Innovation Office of Hungary,
via the projects: 
Nanoplasmonic Laser Inertial Fusion Research Laboratory in the past years
numbered as:
(NKFIH-874-2/2020),
(NKFIH-468-3/2021), 
(2022-2.1.1-NL-2022-00002), 
furthermore by
Optimized nanoplasmonics (K116362), and
Ultrafast physical processes in atoms, molecules, nanostructures 
and biological systems (EFOP-3.6.2-16-2017-00005).


\end{document}